\documentclass[12pt,a4paper]{article}
\usepackage{verbatim}
\usepackage[latin1]{inputenc}
\usepackage[all]{xy} 
\usepackage{graphicx} 
\usepackage[tight]{subfigure}
\usepackage{hyperref}
\usepackage[usenames,dvipsnames,svgnames,table]{xcolor}
\hypersetup{colorlinks=true, citecolor=Blue, linkcolor=Blue, urlcolor=Blue}
\usepackage{xcolor}
\usepackage{cancel}
\usepackage{lineno}

\usepackage{amsmath,amsfonts,amssymb,amsthm,epsfig,epstopdf,url,array}

\usepackage{todonotes}
\usepackage{graphicx}
\usepackage{color}

\def\be{\begin{equation}}
	\def\ee{\end{equation}}
\def\bi{\begin{itemize}}
	\def\ei{\end{itemize}}
\def\ed{\end{document}}

\newcommand{\R}{{\mathbb R}}

\renewcommand{\u}{{\bf u}}

\def\be{\begin{equation}}
\def\ee{\end{equation}}

\theoremstyle{definition}
\newtheorem{defn}{Definition}[section]

\newtheorem{thm}{Theorem}[section]
\newtheorem{remark}{Remark}[section]

\newcommand{\q}{\quad}

\renewcommand{\u}{{\bf u}}

\begin{document}

\title{Energy stability of plane Couette and Poiseuille flows and Couette paradox}

\author{Paolo Falsaperla, Giuseppe Mulone \& Carla Perrone}
\date{}

\maketitle


\begin{abstract}
We study the nonlinear stability of plane Couette and Poiseuille flows with the Lyapunov second method by using the classical $L_2$-energy. We prove that the streamwise perturbations are $L_2$-energy stable for any Reynolds number. This contradicts the results of Joseph \cite{Joseph.1966}, Joseph and Carmi \cite{Joseph.Carmi.1969} and Busse \cite{Busse.1972}, and allows us to prove that the critical nonlinear Reynolds numbers are obtained along two-dimensional perturbations, the spanwise perturbations, as Orr \cite{Orr.1907} had supposed. This conclusion combined with recent results by Falsaperla et al. \cite{Falsaperla.Giacobbe.Mulone.2019}  on the stability with respect to tilted rolls, provides a possible solution to the Couette-Sommerfeld paradox.
\end{abstract}


\section{Introduction}
The study of stability and instability of the classical laminar flows of an incompressible fluid has been studied analytically, numerically and with experiments \cite{Couette.1890},  \cite{Poiseuille.1843},  \cite{Reynolds.1883},  \cite{Orr.1907}, \cite{Sommerfeld.1908}, \cite{Squire.1933},  \cite{Joseph.1976}, \cite{Busse.1972}, \cite{Romanov.1973}, \cite{Orszag.1971}, \cite{Barkley.Tuckerman.1999} - \cite{Barkley.Tuckerman.2007}, \cite{Prigent.Gregoire.Chate.Dauchot.2003}, \cite{Falsaperla.Giacobbe.Mulone.2019}. 

This problem is nowadays object of study 
because the transition from lami\-nar flows to instability, turbulence and chaos is not completely understood and there are some discrepancies between the critical Reynolds numbers of linear and nonlinear analysis and the experiments (the so called \textit{Couette-Sommerfeld paradox}).

The classical results are the following:

a) plane Poiseuille flow is \textit{unstable} for ${\rm Re}>5772$ (Orszag \cite{Orszag.1971});

b) plane Couette flow is \textit{linearly stable} \textit{for any} Reynolds numbers (Romanov \cite{Romanov.1973};

c) in laboratory experiments plane Poiseuille flows undergo transition to three-dimensional turbulence for Reynolds numbers on the order of 1000. In the case of plane Couette flow the lowest Reynolds numbers at which turbulence can be produced and sustained have been shown to be between 300 and 450 both in the numerical simulations and in the experiments \cite{Barkley.Tuckerman.2007}, \cite{Prigent.Gregoire.Chate.Dauchot.2003};

d) nonlinear asymptotic \textit{$L_2$-energy stability} has been proved for Reynolds numbers ${\rm Re}$ below some critical nonlinear value ${\rm Re}_c$ which is of the order $10^2$. In particular Joseph  \cite{Joseph.1966} - \cite{Joseph.1976} proved that ${\rm Re_c} ={\rm Re}^y=20.65$ (and ${\rm Re}^x=44.3$) for plane Couette flow, 
and Joseph and Carmi \cite{Joseph.Carmi.1969} proved that ${\rm Re_c} ={\rm Re}^y=49.55$ (and ${\rm Re}^x=87.6$) for plane Poiseuille flow. 
Here and in what follows ${\rm  Re}^y$ refers to critical value for streamwise (or longitudinal) perturbations, ${\rm Re}^x$ refers to the critical value for spanwise (or transverse) perturbations.	

Moffatt \cite{Moffatt.1990} proved the stability of the basic motion with respect to two-dimensional streamwise perturbations for any Reynolds number, for the case of inhomogeneous perturbation flow of a particular type (see \cite{Moffatt.1990}, pp. 250-252) and $y\in [0, +\infty)$, with a hypothesis about  pressure that vanishes at infinity. Even if he doesn't say so explicitly, he used a weighted energy argument.

The use of \textit{weighted $L_2$-energy} has been fruitful for studying  nonlinear stability in fluid mechanics (see Straughan \cite{Straughan.2004}).

Rionero and Mulone \cite{Rionero.Mulone.1991} studied the nonlinear stability of parallel shear flows with the Lyapunov method in the (ideal) case of stress-free boundary conditions. By using a weighted energy they proved that plane Couette flows and plane Poiseuille flows are conditionally asymptotically stable for any Reynolds numbers. 	

Kaiser et al. \cite{Kaiser.Tilgner.vonWhal.2005} wrote the velocity field in terms of poloidal, toroidal and the mean field components. They used a generalized energy functional ${\cal E}$  (with some coupling parameters chosen in an optimal way) for plane Couette flow, providing conditional nonlinear stability for Reynolds numbers ${\rm Re}$ below ${\rm Re_{\cal E}} = 44.3$, which is larger than the ordinary energy stability limit (it is the energy limit for spanwise perturbations). The method allows the explicit calculation of so-called stability balls in the ${\cal E}$-norm; i.e., the system is stable with respect to any perturbation with ${\cal E}$-norm in this ball.

Kaiser and Mulone \cite{Kaiser.Mulone.2005} proved conditional nonlinear stability  for \textit{arbitrary plane parallel shear flows} up to some value ${\rm Re}_E$ which depends on the shear profile. They used a generalized (weighted) functional $ E $ and proved that ${\rm Re}_E$ turns out to be ${\rm Re}^x$, the ordinary energy stability limit for perturbations independent of $y$ (spanwise perturbations). In the case of the experimentally important profiles, viz. linear combinations of Couette and Poiseuille flow, this number is at least $87.6$, the value for pure Poiseuille flow. For Couette flow it is at least $44.3$

The problem of finding the best conditions for global nonlinear energy stability with respect to three-dimensional perturbations is still an open problem. 
This problem is equivalent to finding the maximum of a functional ratio that arises from the Reynolds energy equation \cite{Reynolds.1883}.

In the nonlinear case it is often assumed that the least stabilizing perturbations, as in the linear case, are the two-dimensional spanwise perturbations, see Orr \cite{Orr.1907}. However, Joseph in his paper on Couette flow \cite{Joseph.1966} proved that the least stabilizing perturbations are the streamwise perturbations and concluded that the Orr result is wrong. Joseph and Carmi \cite{Joseph.Carmi.1969} and Busse \cite{Busse.1972} obtained a similar result of \cite{Joseph.1966} for Poiseuille flow. Also our numerical calculations,  with the Chebyshev polynomials method (see Figure 1), show that the mi\-nimum Reynolds number for the energy method is obtained with respect to streamwise perturbations. Therefore, we  obtain the same numerical results of Joseph for Couette case and Joseph an Carmi and Busse for Poiseuille case (${\rm Re}_c={\rm Re}^y =20.6$ in Couette case, and ${\rm Re}_c={\rm Re}^y=49.55$ in Poiseuille case). 

Despite of the results of Joseph \cite{Joseph.1966}, \cite{Joseph.1976}, Joseph and Carmi \cite{Joseph.Carmi.1969}, and Busse \cite{Busse.1972}, we shall prove that \textit{the conclusion of Joseph} \cite{Joseph.1976} {\it is not correct}.

The first aim of this paper is to prove that streamwise perturbations are nonlinearly exponentially stable for any Reynolds number (i.e. Re$^y=+\infty$). We shall prove this, in Sec. 3, in two ways: we first use a weighted energy with a positive coupling parameter, then we use the classical $L_2$-energy norm.
Although some authors \cite{Reddy.Schmid.Baggett.Henningson.1998} assume that streamwise perturbations are always stable, as far as we know, in the literature there is no precise proof apart from Moffatt's proof in a semi-space \cite{Moffatt.1990}. 

We then observe that  \textit{there is a contradiction} between this result ( Re$^y=+\infty$) and the numerical ones. The  critical Reynolds number is obtained from the maximum problem (see Sec. 3) on the streamwise perturbations: Re$^y$ is a finite real value (Re$^y$= 20.6 for Couette and  Re$^y$=49.5 for Poiseuille).
In Sect. 3 we \textit{suggest how to solve this contradiction through a conjecture}: the maximum is obtained in a subspace of the space of kinematically admissible perturbations, the space of \textit{physically admissible perturbations} competing for the maximum. In this way, we are able to prove that the maximum is reached on two-dimensional perturbations, the spanwise perturbations, and that the streamwise perturbations are stable for any Reynolds number.

This result (i.e. the least stabilizing perturbations are two-dimensional perturbations) justifies the study of the stability with respect to two-dimensional rolls done by Falsaperla et al. \cite{Falsaperla.Giacobbe.Mulone.2019}. These rolls appear at the onset of turbulence in the experiments and so we have a possible explanation of the Couette-Sommerfeld paradox.

The plan of the paper is the following. 

In Sec. 2 we write the non-dimensional perturbation equations of laminar flows between two horizontal rigid planes, and we recall the classical linear stability/instability results. 

In Sec. 3 we recall the classical energy stability results of Orr \cite{Orr.1907}, Joseph \cite{Joseph.1966}, Joseph and Carmi \cite{Joseph.Carmi.1969}, Busse \cite{Busse.1972} (see Drazin and Reid \cite{Drazin.Reid.2004}). Then, we prove nonlinear exponential stability of streamwise perturbations for any Reynolds number, i.e. Re$^y = +\infty$ with two energies: a weighted and a classical $L_2$-energy norm. Then, we study the maximum problem arising from the Reynolds energy \cite{Reynolds.1883}, we arrive at a contradiction and suggest a conjecture to solve it by introducing the space of physical admissible perturbations for the maximum problem. In this space we find the optimal perturbations which give the critical Reynolds number: the spanwise perturbations, as Orr \cite{Orr.1907} had supposed.

In Sec. 4  we recall recent results of Falsaperla et al. \cite{Falsaperla.Giacobbe.Mulone.2019} and give sufficient nonlinear stability conditions of the plane Couette and Poiseuille flows with respect to tilted perturbations (rolls) of an angle $\theta$ with respect to the direction of the motion and fixed wavelength $\lambda$. We prove that they are non-linearly exponentially stable for any Reynolds number less that $\dfrac{{\rm Re}_{Orr}(2\pi /(\lambda\sin \theta))}{\sin\theta}$ where  ${\rm Re}_{Orr}$ is the critical Reynolds number for spanwise perturbations evaluated at $2\pi /(\lambda\sin \theta)$.

In Sec. 5  we  make some final comments.

\section{Laminar flows between two parallel planes}
Given  a reference frame $Oxyz$, with unit vectors ${\bf i},{\bf j}, {\bf k}$, consider the layer $\mathcal D = \R^2 \times [-1, 1]$  of thickness $2$ with horizontal coordinates $x,y$ and vertical coordinate $z$.

Plane parallel shear flows are solutions of the stationary Navier-Stokes equations
\begin{eqnarray}\label{Couette-gen}
	\left\{ \begin{array}{l}
	{\bf U}\!\cdot\!\nabla {\bf U} ={\rm Re}^{-1} \Delta {\bf U}-\nabla P\\[5pt]
	\nabla \cdot {\bf U}=0 ,\\
	\end{array}  \right.
\end{eqnarray}
characterized by the functional form
\be{\bf U}= \begin{pmatrix} \label{basic}
	
	f(z)\\
	0 \\
	0
\end{pmatrix} =  f(z) {\bf i},
\ee
where ${\bf U}$ is the velocity field and $P$ the pressure field, and ${\rm Re}$ is the Reynolds number. 
The function $f(z) : [-1, 1] \to \mathbb R$ is assumed to be sufficiently smooth and is called the shear profile. All the variables are written in a non-dimensional form. To non-dimensionalize the equations and the gap of the layer we use a Reynolds number based on the average shear and half gap $d$ (see \cite{Falsaperla.Giacobbe.Mulone.2019}).

In particular, for fixed velocity at the boundaries $z=\pm 1$, we have the well known profiles:

a) Couette  $f(z)=z$,

b) Poiseuille $f (z) = 1-z^2$.	

\subsection{Perturbation equations}
The perturbation equations to the plane parallel shear flows, in non-dimensional form, are 
\begin{eqnarray}\label{Couette-gen}
	\left\{ \begin{array}{l}
		u_t = -  {\bf u}\!\cdot\!\nabla u+ {\rm Re}^{-1} \Delta u -  (f u_x+f' w)- \dfrac{\partial p}{\partial x}\\[5pt]
		v_t = -  {\bf u}\!\cdot\!\nabla v+ {\rm Re}^{-1} \Delta v -  f v_x - \dfrac{\partial p}{\partial y}\\[5pt]
		w_t = -  {\bf u}\!\cdot\!\nabla w+  {\rm Re}^{-1}\Delta w -  f w_x- \dfrac{\partial p}{\partial z}\\	[5pt]	
		\nabla \cdot\u=0 ,\\
	\end{array}  \right.
\end{eqnarray}
where ${\bf u}= u{\bf i}+ v {\bf j}+w{\bf k}$ is the perturbation to the velocity field, $p$ is the perturbation to the pressure field.

Throughout the paper, we use the symbol $h_x$ as $\frac {\partial h} {\partial x}$, $h_t$ as $\frac {\partial h} {\partial t}$, etc., for any function $h$.

To system (\ref{Couette-gen}) we append the rigid boundary conditions $${\bf u}(x,y,\pm 1,t)=0, \q (x,y,t) \in  \R^2 \times (0, +\infty), $$ and the initial condition

$${\bf u}(x,y,z,0)= {\bf u_0}(x,y,z), \q {\rm in}  \q\mathcal D,$$
with ${\bf u_0}(x,y,z)$ solenoidal vector which vanishes at the boundaries.

\begin{defn}
	\textit{We define streamwise (or longitudinal) perturbations the perturbations ${\bf u}, p$  which do not depend on $x$. 
}\end{defn}

\begin{defn}
	\textit{	We define spanwise (or transverse) perturbations the perturbations ${\bf u}, p$  which do not depend on $y$.} 
\end{defn}


\subsection{Linear stability/instability}

Assume that both ${\bf u}$ and $\nabla p $ are $x,y$-periodic with periods $a$ and $ b$ in the $x$ and $y$ directions, respectively,  with wave numbers $( a, b) \in \R^2_+$  . In the following it suffices therefore to consider functions over the periodicity cell 
$$\Omega= [0, \frac{2\pi}{a}]\times [0, \frac{2\pi}{ b}] \times [- 1, 1] .$$

As the basic function space, we take $L_2(\Omega)$, which is the space of square-summable functions in $\Omega$ with the scalar product denoted by

$$(g,h) = \int_0^{\frac{2\pi}{a}} \int_0^{\frac{2\pi}{ b}} \int_{-1}^1 g(x,y,z) h(x,y,z) dxdydz, $$ 

and the norm given by

$$\Vert g \Vert = \Bigl[\int_0^{\frac{2\pi}{a}} \int_0^{\frac{2\pi}{ b}} \int_{-1}^1 g^2(x,y,z) dxdydz\Bigr]^{\frac{1}{2}}.$$ 

We recall that the classical results of Romanov \cite{Romanov.1973}  prove that Couette flow is \textit{linearly stable} for \textit{any Reynolds} number. Instead, Poiseuille flow is unstable for any Reynolds number bigger that $5772$ (Orszag \cite{Orszag.1971}).

The linear stability/instability is obtained by studying the linearised system  by neglecting the terms ${\bf u}\!\cdot\!\nabla u$, ${\bf u}\!\cdot\!\nabla v$, ${\bf u}\!\cdot\!\nabla w$ in \eqref{Couette-gen}.

We observe that, in the linear case, the Squire theorem \cite{Squire.1933} holds and the most destabilizing perturbations are two-dimensional, in particular the spanwise  perturbations (see Drazin and Reid \cite{Drazin.Reid.2004}, p. 155). The critical  Reynolds value, for Poiseuille flow, can be obtained by solving the celebrated Orr-Sommerfeld equation (see Drazin and Reid \cite{Drazin.Reid.2004}).

\section{Nonlinear energy stability}

Here we  study \textit{ the nonlinear energy stability with the Lyapunov method}, by using the classical energy 

$$V(t) = \dfrac{1}{2}[\Vert u \Vert^2 +  \Vert v \Vert^2 + \Vert w \Vert^2 ]. $$
We obtain \textit{sufficient conditions of global nonlinear stability}.

Taking into account the solenoidality of ${\bf u}$ and the boundary condition, we  write the energy identity \cite{Reynolds.1883}

\be\label{Energy}
\dot V= -(f'w,u) - {\rm Re}^{-1} [\Vert \nabla u \Vert^2+\Vert \nabla v \Vert^2+\Vert \nabla w \Vert^2], 
\ee
and we have 
\be\label{time-en}
\begin{array}{l}
	\dot V= -(f'w,u) - {\rm Re}^{-1} [\Vert \nabla u \Vert^2+\Vert \nabla v \Vert^2+\Vert \nabla w \Vert^2] =\\[3mm]
	= \left(\dfrac{-(f'w,u)}{\Vert \nabla u \Vert^2+\Vert \nabla v \Vert^2+\Vert \nabla w \Vert^2} - \dfrac{1}{{\rm Re}}\right)\Vert \nabla {\bf u} \Vert^2  \le \\[3mm]
	\le \left( \dfrac{1}{\rm Re}_c - \dfrac{1}{{\rm Re}}\right)\Vert \nabla {\bf u} \Vert^2 ,
\end{array}\ee 
where 
\be\label{maxRe}\dfrac{1}{\rm Re}_c=m=  \max_{\cal S} \dfrac{-(f'w,u)}{\Vert \nabla u \Vert^2+\Vert \nabla v\Vert^2+\Vert \nabla w \Vert^2}, \ee
and $\cal S$ is the space of the \textit{kinematically admissible fields } 
\be \begin{array}{l}\label{spaceS}
	{\cal S}= \{u, v, w \in H^2 (\Omega), \; u=v=w=w_z=0 \hbox{ on the boundaries,}\\[3mm] \hbox{ periodic in \textit{x}, and \textit{y},}  \q u_x+v_y+w_z=0,\q  \Vert \nabla {\bf u} \Vert>0\},
\end{array}
\ee
and $ H^2(\Omega)$ is the Sobolev space of the functions which are in $L_2(\Omega)$ together with their first and second generalized derivatives.

The Euler-Lagrange equations of this maximum problem are given by 

\be\label{EL-Orr0}
(f' w {\bf i}+ f' u {\bf k}) - 2m \Delta {\bf u} = \nabla \lambda,
\ee
where $\lambda$ is a Lagrange multiplier.  Assuming as Orr \cite{Orr.1907} that the maximum is achieved for spanwise perturbations, $\dfrac{\partial}{\partial y}\equiv 0$,   by taking the third  component of the \textit{double-curl} of  (\ref{EL-Orr0}) and by using the solenoidality condition $u_x+w_z=0$ and the boundary conditions, 
we obtain the \textit{Orr equation} \cite{Orr.1907}
\be \label{Orr-eq}
\dfrac{{\rm Re}_c}{2} (f'' w_x+2f' w_{xz})+ \Delta \Delta w=0.	
\ee	

By solving this equation  we obtain the Orr results: for Couette and Poiseuille flows, we have ${\rm Re}_{Orr}= {\rm Re}_c= 44.3$ (cf. Orr \cite{Orr.1907} p. 128) and ${\rm Re}_{Orr}={\rm Re}_c= 87.6$ (cf. Drazin and Reid \cite{Drazin.Reid.2004} p. 163), respectively (the critical values are converted to the  dimensionless form we have used here).

These nonlinear stability conditions have been obtained by Orr \cite{Orr.1907} in a celebrated paper, by using the Reynolds \cite{Reynolds.1883} energy \eqref{Energy} (see Orr \cite{Orr.1907} p. 122).

In his paper Orr \cite{Orr.1907} writes: ``\textit{Analogy with other problems leads us to assume that disturbances in two dimensions will be less stable than those in three; this view is confirmed by the corresponding result in case viscosity is neglected}". He also says:  ``\textit{The three-dimensioned case was attempted, but it proved too difficult}".

Orr considers spanwise perturbations ($\frac{\partial}{\partial y}\equiv 0$, he also considers two-dimen\-sional perturbations, $v\equiv 0$), the same perturbations that are utilized in the linear case by using Squire transformation (cf. Squire \cite{Squire.1933}, Drazin and Reid \cite{Drazin.Reid.2004}).
The critical value he finds, in the Couette case,  is Re$^{x}=44.3$, where Re$^{x}$ \textit{is the critical Reynolds number with respect to spanwise perturbations} (see Orr \cite{Orr.1907} p. 128, Joseph \cite{Joseph.1976} p. 181).

Joseph in his monograph \cite{Joseph.1976}, p. 181, says: ``\textit{Orr's assumption about the form of the disturbance which increases at the smallest {\rm Re} is not correct since we shall see that the energy of an $x$-independent disturbance (streamwise perturbations) can increase when  ${\rm Re} >2 \sqrt{1708}\simeq 82.65$}" (in our dimensionless form Re$^y$=20.6)

Joseph also gives a Table of values of the principal eigenvalues (critical ener\-gy Reynolds numbers) which depend on  a parameter  $\tau$ which varies from $0$ (streamwise perturbations) to $1$ (spanwise perturbations) and concludes that \textit{the value ${\rm Re}^y = 20.6$, 
is the limit for energy stability when $\tau=a=0$ (streamwise perturbations)}, where $a$ is the wave number in the $x$-direction.

Busse (\cite{Busse.1972}, p. 29) in his paper writes 
``\textit{Numerical computations suggest that the eigenvalue $R_E$ is attained for $x$-independent solutions. Since this result has contradicted the physical intuition of earlier investigators in this field, it is desirable to find a rigorous proof for this property}". However he remarks in a note on p. 29:  ``\textit{Joseph first found that the minimizing solution in the case of Couette flow was independent of $x$. He also gave a proof of this fact. A gap of his proof has been found, however, recently by J. Serrin (private communication by D. D. Joseph) who pointed out that Joseph did not account for the possibility that the required minimum could appear at the end point $k_x=k$}".  

Drazin and Reid in their monograph \cite{Drazin.Reid.2004}, p. 430, note:  ``\textit{The determination of the least eigenvalue of equations (53.21)} (in our case,  ${\rm Re}_c=1/m$ in equation \eqref{EL-Orr0}) \textit{is clearly a formidable problem in general and results are known for only a few cases}".
They cite the results of Joseph \cite{Joseph.1966} for Couette flow, Joseph and Carmi \cite{Joseph.Carmi.1969} and Busse \cite{Busse.1972} for Poiseuille flow. They also say (\cite{Drazin.Reid.2004}, p. 430):  ``\textit{the least eigenvalue ... is still associated with a two dimensional disturbance but one which varies only in the $yz$-plane, i.e. the perturbed flow consists of rolls whose axes are in the directions of the basic flow}" (streamwise perturbations).

We note that  our numerical calculations,  with the Chebyshev polynomials method (see Figure 1), also show that the minimum Reynolds number for the energy method is obtained with respect to streamwise perturbations. Therefore, we  obtain the same nume\-rical results of Joseph for Couette case, Joseph an Carmi and Busse for Poiseuille case (${\rm Re}_c={\rm Re}^y =20.6$ in Couette case, and ${\rm Re}_c={\rm Re}^y=49.5$ in Poiseuille case). 


Despite of the results of Joseph \cite{Joseph.1966}, \cite{Joseph.1976}, Joseph and Carmi \cite{Joseph.Carmi.1969}, Busse \cite{Busse.1972} and the observations of Drazin and Reid \cite{Drazin.Reid.2004}, we shall prove below that the numerical calculations of Joseph are correct, however \textit{the conclusion of Joseph} \cite{Joseph.1976} {\it is not correct}. In fact, in the classical $L_2$-energy, the \textit{streamwise perturbations are always stable}. This result for the first time has been proved by Moffatt \cite{Moffatt.1990} for the case of inhomogeneous perturbation flow of a particular type (see \cite{Moffatt.1990}, pp. 250-252) and $y\in [0, +\infty)$, with a hypothesis about  pressure that vanishes at infinity.

In the next subsection we prove that the streamwise perturbations are stable for any Reynolds number. This means that the result of Orr is correct (see \cite{Falsaperla.Giacobbe.Mulone.2019}, \cite{Falsaperla.Mulone.Perrone.2021}) as Serrin had also observed \cite{Busse.1972}. Moreover, in what follows, we make a conjecture and  we prove that the least stabilizing \textit{physical  perturbations} competing for maximum are two-dimensional, and they are the spanwise.

\subsection{Stability of streamwise perturbations for any Reynolds number}

We assume that the perturbations are \textit{streamwise}, i.e. they \textit{do not depend on} $x$ ($\frac{\partial}{\partial x} \equiv 0$).

Therefore the perturbation equations \eqref{Couette-gen}  become

\begin{eqnarray}\label{Couette-long}
	\left\{ \begin{array}{l}
		u_t = -  {\bf u}\!\cdot\!\nabla u+  {\rm Re}^{-1}\Delta u -  f' w\\	[5pt]	
		v_t = -  {\bf u}\!\cdot\!\nabla v+  {\rm Re}^{-1}\Delta v  - \dfrac{\partial p}{\partial y}\\	[5pt]	
		w_t = -  {\bf u}\!\cdot\!\nabla w+ {\rm Re}^{-1} \Delta w- \dfrac{\partial p}{\partial z}\\	[5pt]			
		v_y+w_z=0  .\\
	\end{array}  \right.
\end{eqnarray}

i) {\bf Weighted energy}\\
\\
We define a \textit{weighted energy} (Lyapunov function) equivalent to the classical energy norm and show that the streamwise perturbations cannot destabilize the basic Couette or Poiseuille flows.

First we  introduce an  arbitrary positive number  $\beta$. Then, we multiply (\ref{Couette-long})$_1$ by $\beta u$ and integrate over $\Omega$. Besides, we multiply (\ref{Couette-long})$_2$ and  (\ref{Couette-long})$_3$ by $v$ and $w$ and integrate over $\Omega$. 
By taking into account the solenoidality of ${\bf u}$, the boundary conditions and the periodicity, we have

$$\dfrac{d}{dt} [\dfrac{\beta \Vert u\Vert^2}{2}]= -\beta (f' w, u) - \beta {\rm Re}^{-1}\Vert \nabla u \Vert^2$$

$$\dfrac{d}{dt} [\dfrac{\Vert v\Vert^2+ \Vert w\Vert^2}{2}]=  - {\rm Re}^{-1} (\Vert \nabla v \Vert^2+ \Vert \nabla w )\Vert^2 .$$

By using the arithmetic-geometric mean inequality, we have

$$-\beta(f' w,u) \le \beta \dfrac{M^2}{2\epsilon} \Vert w \Vert^2 + \beta \dfrac{\epsilon}{2} \Vert u \Vert^2, $$

where $M= \max_{[-1,1]} \vert f'(z)\vert$ and $\epsilon$ is an arbitrary positive number to be chosen.

Now we define  the Lyapunov function (weighted energy norm)

\be\label{energy-beta} E(t)= \dfrac{1}{2}[\beta \Vert u \Vert^2 +  \Vert v \Vert^2 + \Vert w \Vert^2 ], \ee

and choose $ \epsilon = \dfrac{\pi^2}{{\rm 4Re}}.$ From the above inequality and the use of the Poincar\'{e} inequality $\dfrac{\pi^2}{4}\Vert g \Vert^2 \le \Vert \nabla g \Vert^2$ ($g=u$, $g=v$, $g=w$), we have

$$\dot E \le \dfrac{1}{2}[(\beta \dfrac{4M^2 {\rm Re}}{\pi^2}- \dfrac{\pi^2}{{2\rm Re}})\Vert w\Vert^2 - \dfrac{\pi^2}{{2\rm Re}} \Vert v \Vert^2 - \beta \dfrac{\pi^2}{4\rm Re} \Vert u \Vert^2].$$

By choosing $\beta = \dfrac{\pi^4}{16M^2\rm Re^2},$ we finally have

$$ \dot E \le - \dfrac{\pi^2}{{4\rm Re}} E.$$

Integrating this inequality, we have 
the exponential decay

\be\label{expdecay} E(t) \le E(0) \exp\{-\dfrac{\pi^2}{{4\rm Re}} t\}.\ee

This inequality implies global nonlinear exponential stability of the basic Couette or Poiseuille flows with respect to  the streamwise perturbations for any Reynolds number.\\
\\
ii) {\bf Classical $L_2$-energy}\\
\\
Now we use  the {\textit {classical energy norm}} and show that the streamwise perturbations cannot destabilize the basic Couette or Poiseuille flows.

 We multiply (\ref{Couette-long})$_1$ by $u$ and integrate over $\Omega$. Besides, we multiply (\ref{Couette-long})$_2$ and  (\ref{Couette-long})$_3$ by $v$ and $w$ and integrate over $\Omega$. 
By taking into account of the solenoidality of ${\bf u}$, the boundary conditions and the periodicity, as before we have

$$\dfrac{d}{dt}\dfrac{\Vert u \Vert ^2}{2}= - (f' u, w) - {\rm Re}^{-1}\Vert \nabla u \Vert ^2, $$

$$ \dfrac{d}{dt} (\dfrac{\Vert v \Vert ^2}{2} + \dfrac{\Vert w \Vert ^2}{2}) = -{\rm Re}^{-1} [ \Vert \nabla v \Vert ^2 + \Vert \nabla w \Vert ^2] .$$

By using the Poincar\'e inequality and integrating last equation, we have:

\be 
\begin{split} 
	& \dfrac{d}{dt}(\dfrac{\Vert v \Vert ^2}{2} + \dfrac{\Vert w \Vert ^2}{2})  = -{\rm Re}^{-1} [ \Vert \nabla v \Vert ^2 + \Vert \nabla w \Vert ^2] \leq -C  (\Vert v \Vert ^2 + \Vert w \Vert ^2)\\
	& \Rightarrow \Vert v \Vert ^2 + \Vert w \Vert ^2\leq H_0 e^{-2Ct}, \quad  C= \dfrac{\pi^2}{4 {\rm Re}}, \quad H_0= \Vert v_0 \Vert ^2 + \Vert w_0 \Vert ^2.
\end{split}
\label{energiavw}
\ee

Now we consider the equation depending on $u$ and define $M= \max_{[-1,1]} \vert f'(z)\vert$. We have the following inequalities:

\be
\begin{split} 
\dfrac{d}{dt}\dfrac{\Vert u \Vert ^2}{2} & = - (f' u, w) -  {\rm Re}^{-1} \Vert \nabla u \Vert ^2 \leq { M \Vert u \Vert \Vert w \Vert - \rm Re}^{-1} \Vert \nabla u \Vert ^2 \leq \\
& \leq  M (\dfrac{\Vert u \Vert ^2}{2 \epsilon} + \dfrac{\epsilon}{2} {\Vert w \Vert ^2}) - {\rm Re}^{-1} \Vert \nabla u \Vert ^2 \leq  M (\dfrac{\Vert u \Vert ^2}{2 \epsilon} + \dfrac{\epsilon}{2} {\Vert w \Vert ^2}) - \\
& - C \Vert u \Vert ^2  = (\dfrac{M}{2 \epsilon}-C) \Vert u \Vert ^2 + \dfrac{\epsilon}{2}  M \Vert w \Vert ^2 =  -\dfrac{C}{2} \Vert u \Vert ^2 + \dfrac{M^2}{C} \dfrac{\Vert w \Vert ^2}{2}  
\end{split}
\ee
where $ \epsilon = \dfrac{M}{C}$, and $C=\dfrac{\pi^2}{4 {\rm Re}}.$

We use this inequality and \eqref{energiavw} to obtain
\be
\begin{split}
\dfrac{d}{dt}\Vert u \Vert ^2 & \leq -C \Vert u \Vert ^2 + \dfrac{ M^2}{C} \Vert w \Vert ^2  \leq -C \Vert u \Vert ^2 + \dfrac{M^2}{C} (\Vert v \Vert ^2 +  \Vert w \Vert ^2) \leq \\
	& \leq -C \Vert u \Vert ^2 + \dfrac{M^2}{C} H_0 e^{-2Ct}.
\end{split}
\ee

Integrating last inequality, we have
\be\label{int-u}
\Vert u \Vert ^2 \leq e^{-Ct} [k - \dfrac{M^2}{C^2}H_0e^{-Ct}] = k e^{-Ct} - \dfrac{M^2}{C^2}H_0 e^{-2Ct},
\ee
with $k=K_0 + \dfrac{M^2}{C^2} H_0, \quad K_0=\Vert u_0 \Vert ^2$.
%

We introduce the classical energy

\be\label{energy-beta} L(t)= \dfrac{1}{2}[\Vert u \Vert^2 +  \Vert v \Vert^2 + \Vert w \Vert^2 ], \ee

and observe that the initial energy is given by  $L_0=\dfrac{H_0 + K_0}{2}$. Adding the (\ref{energiavw}) and the (\ref{int-u}) we finally have: 

\be\label{L-en}
\begin{split}
	L(t) & \leq H_0 e^{-2Ct} + (K_0 + \dfrac{M^2}{C^2} H_0) e^{-Ct} - \dfrac{M^2}{C^2}H_0 e^{-2Ct} \leq\\
	& \leq L_0 e^{-2Ct} + (L_0 + \dfrac{M^2}{C^2} L_0) e^{-Ct}  =\\
	& = L_0 (e^{-2Ct} + e^{-Ct} + \dfrac{M^2}{C^2} e^{-Ct}).
\end{split}
\ee

This inequality implies global nonlinear exponential stability of the basic Couette or Poiseuille flows with respect to  the streamwise perturbations for any Reynolds number.\\
\\
Therefore, using two different approaches, we have proved: 

\begin{thm}
	\textit{Assuming the perturbations to the basic shear flows (\ref{basic}) are streamwise, then we have nonlinear stability according to (\ref{expdecay}) and \eqref{L-en}, both with the weighted and the classical energy. 
} \end{thm}

Denoting by  ${\rm  Re}^y$ the critical Reynolds number with respect to streamwise perturbations, from this Theorem we have 
$ {\rm  Re}^y = +\infty$, i.e., \textit{the streamwise perturbations cannot destabilize the basic flows.}

\begin{remark}
	We note that this result is also in agreement with what observe Reddy et al. \cite{Reddy.Schmid.Baggett.Henningson.1998}:  ``\textit{Linear stability analysis implies that the disturbance energy of sufficiently small perturbations to laminar flow decays to $0$ as $t\to \infty$. But for finite times the energy of disturbances may grow linearly... For a given streamwise and spanwise wave number one can determine a disturbance, called an optimal, which yields the greatest transient linear growth. In channel flows, the optimals which yield the most disturbance growth are independent, or nearly independent, of the streamwise coordinate}".
\end{remark}

\subsection{Nonlinear stability with respect to three-dimensional perturbations}

We study the nonlinear stability with the Lyapunov method, by using the classical energy 

$$V(t) = \dfrac{1}{2}[\Vert u \Vert^2 +  \Vert v \Vert^2 + \Vert w \Vert^2 ]. $$

By writing the energy identity

\be
\dot V= -(f'w,u) - {\rm Re}^{-1} [\Vert \nabla u \Vert^2+\Vert \nabla v \Vert^2+\Vert \nabla w \Vert^2], 
\ee
we have 
\be
\begin{array}{l}
	\dot V= -(f'w,u) - {\rm Re}^{-1} [\Vert \nabla u \Vert^2+\Vert \nabla v \Vert^2+\Vert \nabla w \Vert^2] =\\[3mm]
	= \left(\dfrac{-(f'w,u)}{\Vert \nabla u \Vert^2+\Vert \nabla v \Vert^2+\Vert \nabla w \Vert^2} - \dfrac{1}{{\rm Re}}\right)\Vert \nabla {\bf u} \Vert^2  \le \\[3mm]
	\le\left(m - \dfrac{1}{{\rm Re}}\right)\Vert \nabla {\bf u} \Vert^2 ,
\end{array}\ee 
where 
\be \dfrac{1}{{\rm Re}_c} = m= \max_{\cal S} \dfrac{-(f'w,u)}{\Vert \nabla u \Vert^2+\Vert \nabla v\Vert^2+\Vert \nabla w \Vert^2}, \ee
and $\cal S$ is the space of the \textit{kinematically admissible fields} \eqref{spaceS}.

The Euler-Lagrange equations of this maximum problem are given by 

\be\label{EL-Orr1}
- f' w {\bf i} -  f' u {\bf k} + 2 m \Delta {\bf u} = \nabla \lambda,
\ee
where $\lambda$ is a Lagrange multiplier.

We define 
$$\zeta = v_x - u_y $$ 
(it is linked to the toroidal part of the decomposition of the velocity vector ${\bf u}$ in the poloidal, toroidal and the mean flow, see \cite{Kaiser.Mulone.2005}, \cite{Kaiser.Tilgner.vonWhal.2005})
and take the third component of the \textit{double curl} of (\ref{EL-Orr1}) and the third component of the \textit{curl} of (\ref{EL-Orr1}). We obtain the system of the Euler-Lagrange equations written in terms of  $\zeta$ and $w$: 

\be\label{elwz}
\begin{cases}
	f' (\zeta _y+2w_{xz})+ f''w_x +2 m \Delta \Delta w=0\\
	 f' w_y + 2 m \Delta \zeta =0,
\end{cases}
\ee

with the boundary conditions 
\be\label{elwz-bc0}
w=w_z=0, \; \zeta =0.
\ee

%

The eigenvalue problem \eqref{elwz} - \eqref{elwz-bc0} is solved with the Chebyshev  method by using 80 polynomials. The maximum we obtain corresponds exactly to the critical Reynolds numbers obtained by Joseph \cite{Joseph.1966} and Joseph and Carmi \cite{Joseph.Carmi.1969} and Busse \cite{Busse.1972} (i.e. the minimum Reynolds number is reached for the streamwise perturbations). We report these results in Figure 1.

\begin{figure}[h]{\label{Couette-tridim}}
	\begin{center}
		\includegraphics[width=10cm]{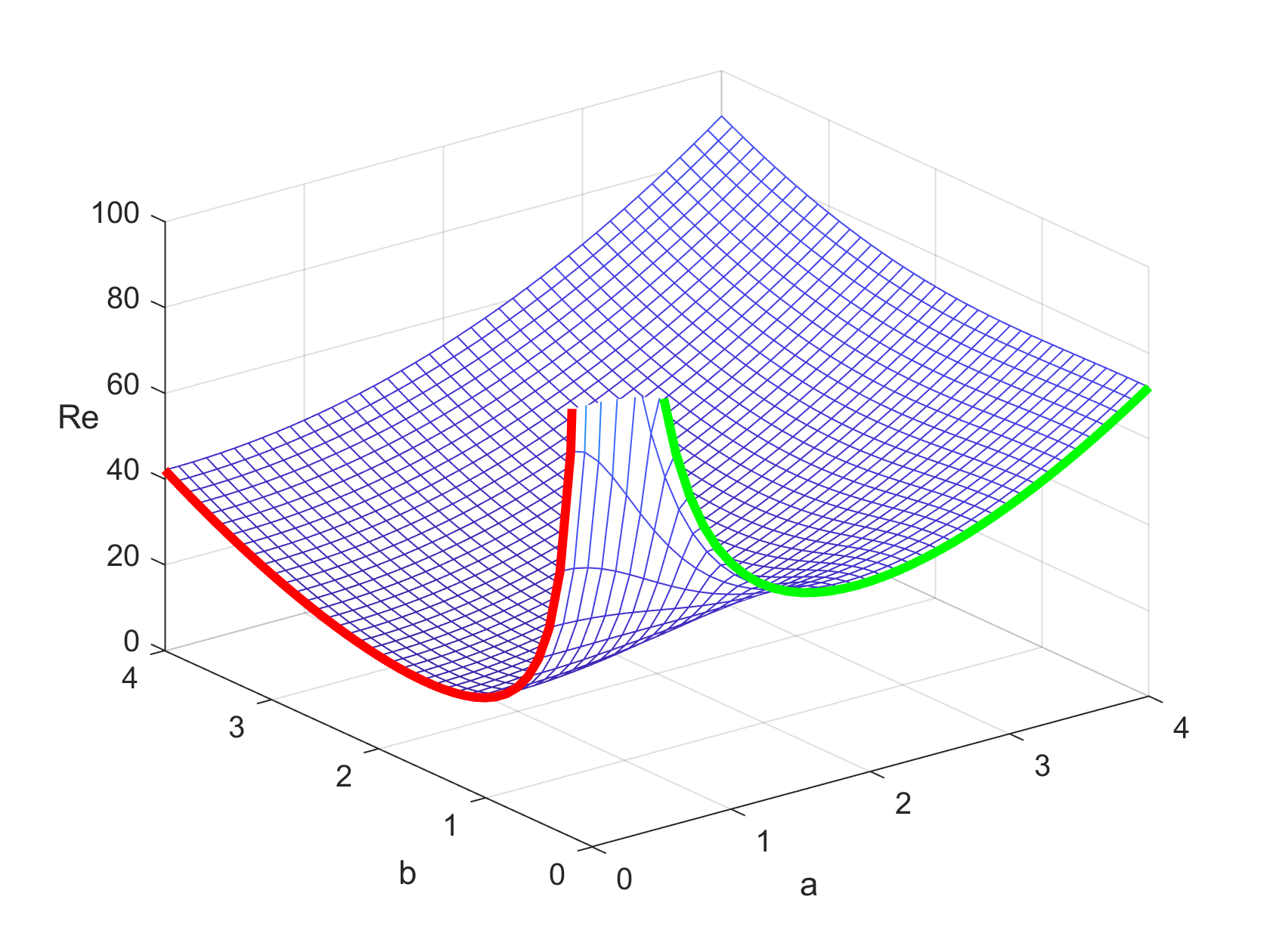}
	\end{center}  \caption{Plane Couette energy Orr-Reynolds  number ${\rm Re}={\rm Re}_c$ as function of the wave numbers $\rm a$ and  $\rm b$,  for system \eqref{elwz} with rigid boundary conditions. The absolute minimum is $\rm R_c=20.6$ and it is achieved for wavenumbers $\rm a=0$, $\rm b=1.6$. For Poiseuille flow a similar picture can be drawn.}
\end{figure}

This conclusion and the results we have obtained in  Subsection 3.1 for streamwise perturbations are in an \textit{obvious contradiction}.  Probably this contradiction is due to the choice of the space of \textit{kinematically admissible} perturbations  where we look for the maximum.  This space is too large and likely contains perturbations which are not admissible as \textit{physical perturbations} competing for the maximum.

\textit{How can we solve this contradiction?}

We first observe that  if the streamwise perturbations (now $w_x=0$ and $\zeta_x=0$) are stable for any Reynolds number, than,  from \eqref{elwz}$_1$,  $m=0$ and this implies that $\zeta_y=0$.
Equation \eqref{elwz}$_2$ implies that also $w_y=0$. If now we consider plan-form perturbations (see Chandrasekhar \cite{Chandrasekhar.1961}, p.24 formula (111))  we have $v= \dfrac{1}{a^2+b^2}[w_{yz}-\zeta_x]=0$. So we are led to speculate:

{\bf Conjecture}

A \textit{possible  answer} to the contradiction is this: we introduce the \textit{subspace 	${\cal S}_0$ of the physical admissible  perturbations}	 which is the subspace of  ${\cal S}$ consisting  of maximizing functions $u, v, w \in {\cal S}$ such that $v=0$ and we  \textit{conjecture} that the maximum $m$ is assumed among the functions of this subspace. Furthermore, we observe that in the numerator of \eqref{maxRe} does not appear explicitly the field $v$.  

With this conjecture, we see immediately that the streamwise perturbations are always stable and the maximum is achieved on the spanwise perturbations.

In fact, we have

\be\label{maxRe-1} m=  \max_{{\cal S}_0} \dfrac{-(f'w,u)}{\Vert \nabla u \Vert^2+\Vert \nabla w \Vert^2}. \ee

The Euler-Lagrange equations of this maximum problem are given by 

\be\label{EL-Orr2}
- f' w {\bf i} -  f' u {\bf k} + 2 m \Delta {\bf u} = \nabla \lambda,
\ee
where $\lambda$ is a Lagrange multiplier, $\nabla \lambda = (\lambda_x, 0, \lambda_z)^T$.

We take the third component of the double curl of (\ref{EL-Orr2}) and the third component of the curl of (\ref{EL-Orr2}). We obtain the system of the Euler-Lagrange equations written in terms of  $\zeta$ and $w$: 

\be\label{elwz-3}
\begin{cases}
	f' (\zeta _y+2w_{xz})+ f''w_x +2 m \Delta \Delta w=0\\
	f' w_y + 2 m \Delta \zeta =0,
\end{cases}
\ee
with the boundary conditions 
\be\label{elwz-bc}
w=w_z= \zeta =0,
\ee
where now $\zeta= - u_y$.

We take the second component of the double curl of (\ref{EL-Orr2}) to get

\be\label{j-comp-el}
f' \zeta_z+f'' \zeta-f'w_{xy}=0.
\ee
From this equation and \eqref{elwz-3}$_2$ we have that $\zeta$ and all its derivative with respect to $z$ are zero on the boundaries. This implies that $\zeta \equiv 0$, hence $u_y=0$ and from \eqref{elwz-3}$_2$ also $w_y=0$. Therefore, $u= u(x,z), v=0, w=w(x,z)$ and the less stabilizing perturbations which satisfy the equation
\be\label{spanwi}
2f' w_{xz}+ f''w_x +2 m \Delta \Delta w=0,
\ee
with boundary conditions $w=w_z=0$ on $z=\pm 1$, are the spanwise perturbations, as Orr \cite{Orr.1907} had supposed.
Moreover, if we consider streamwise perturbations, from the equation
$$m\Vert \Delta w \Vert^2=0 ,$$
 we immediately find $m=0$, i.e. ${\rm Re}^{y} = +\infty$. We report these results in Figure 2 where the critical Reynolds number versus wave numbers for spanwise perturbations are shown.
 

\begin{figure}[h]{\label{Couette-spanwise-inf}}
	\begin{center}
		\includegraphics[width=6.5cm]{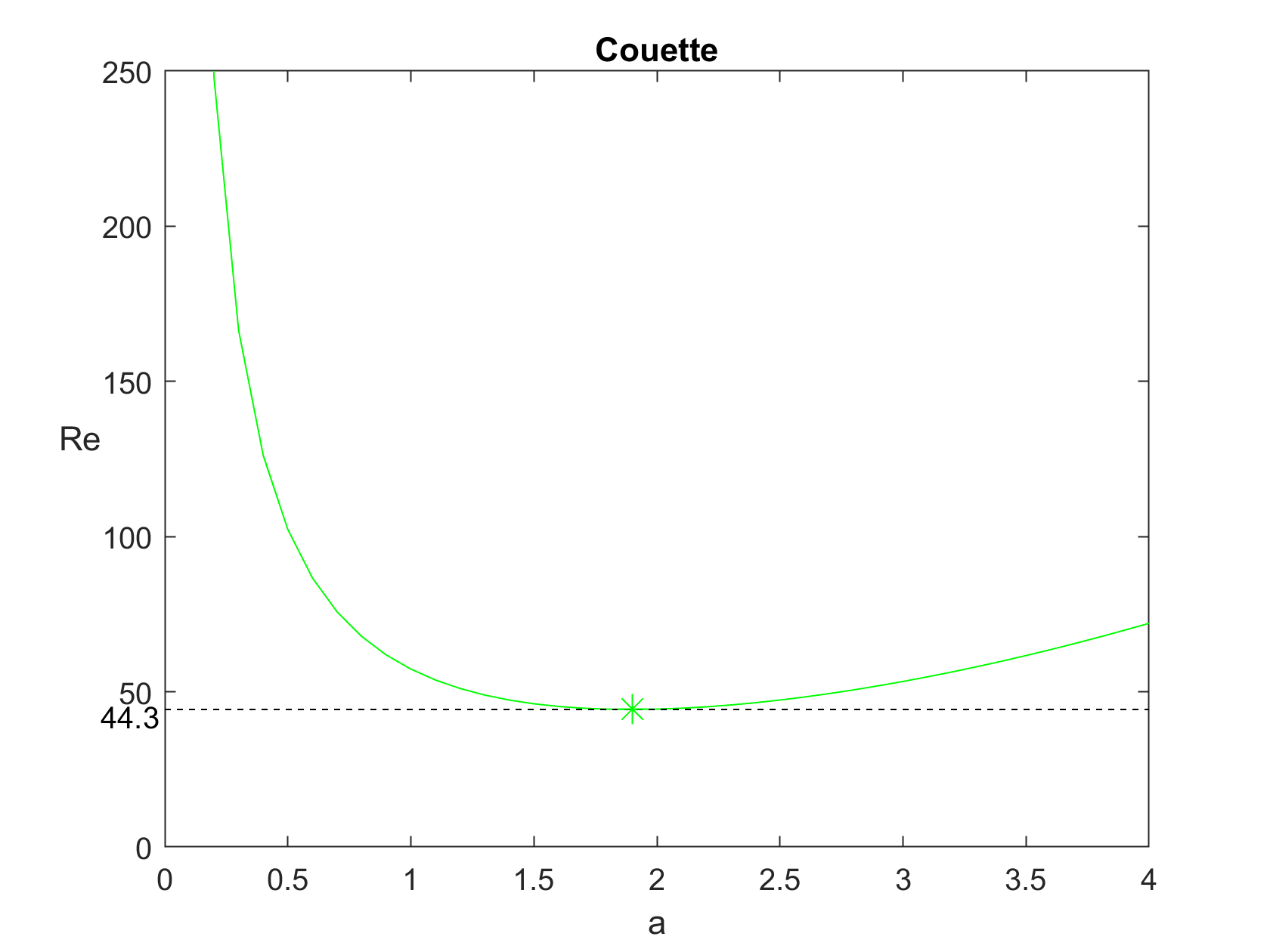}  \quad
	\includegraphics[width=6.5cm]{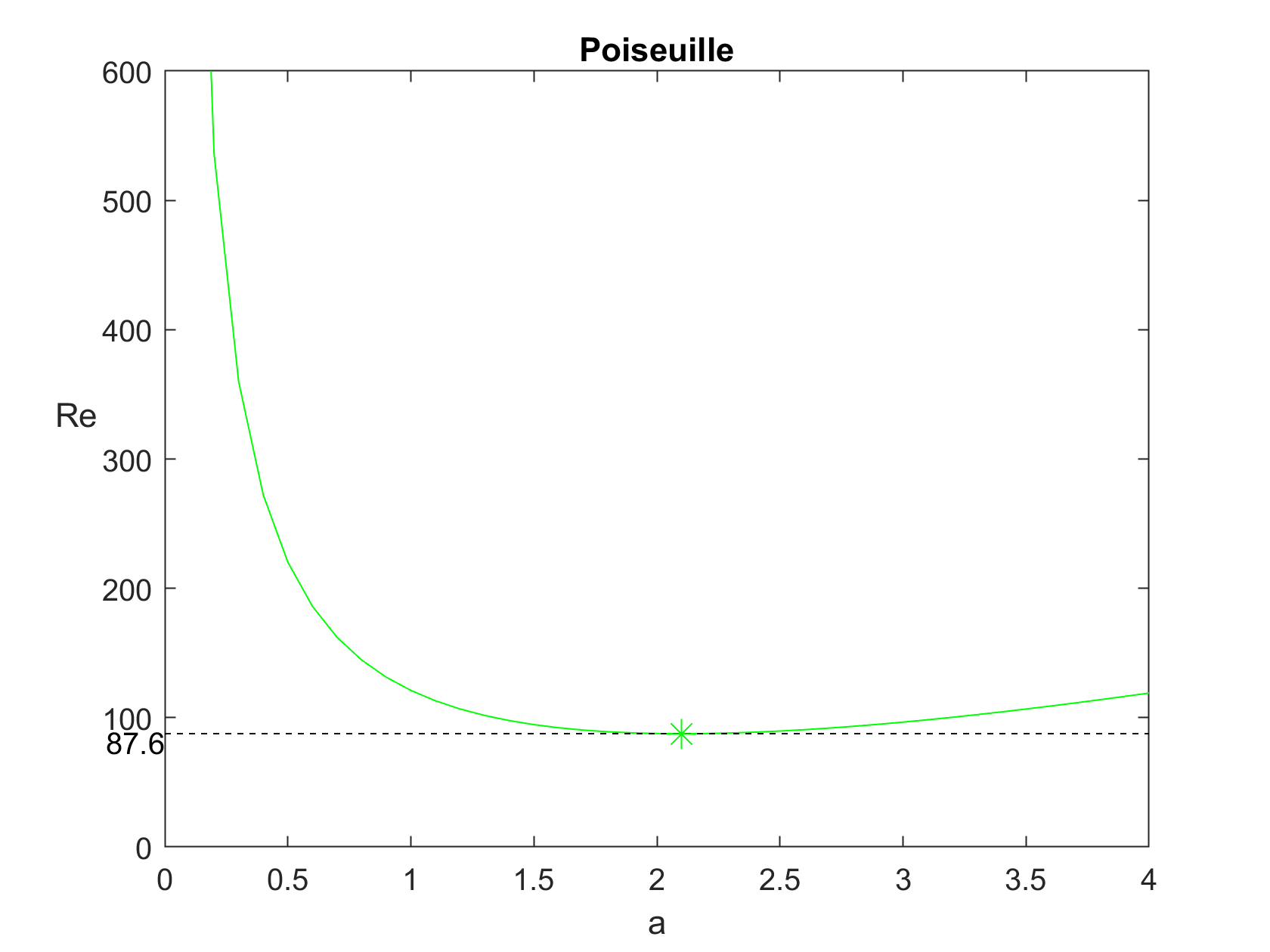}
\end{center}  \caption{Reynolds number versus wave number for spanwise perturbations for plane Couette (left) and   Poiseuille (right) flows.} 
\end{figure}

\vskip 1cm

\section{Two-dimensional tilted perturbations}

We have seen that the least stabilizing perturbations in the linear case are two-dimensional (the Squire \cite{Squire.1933} theorem). In the nonlinear case, we have seen, in the previous section,  that the least stabilizing perturbations are the two-dimensional spanwise perturbations. Since the experiments of Prigent et al. \cite{Prigent.Gregoire.Chate.Dauchot.2003} show that at the onset of instability some tilted perturbations (rolls with axes in the $Oxy$-plane) appear (see Prigent et al. \cite{Prigent.Gregoire.Chate.Dauchot.2003}, Fig. 3), it is obvious and important to study linear and nonlinear energy stability with respect to two-dimensional perturbations and look for those perturbations which first appear at the onset of instability and compare the critical Reynolds numbers 
obtained with the experimental ones. This could solve the Couette-Sommerfeld paradox, because now it is proved that the least stable perturbations at the onset of turbulence are two-dimensional perturbations. For this reason here we briefly recall the main results of \cite{Falsaperla.Giacobbe.Mulone.2019}.

Falsaperla et al. \cite{Falsaperla.Giacobbe.Mulone.2019} considered tilted perturbations (rolls with axes in the plane $Oxy$ ) of an angle $\theta$ with  the $x$-direction, i.e.,  the perturbations ${\bf u}, p$ along the $x'$ axis which forms an angle $\theta\in (0, \frac{\pi}{2}]$ with the direction $x$ and does not depend on $x'$.
For this, they considered an arbitrary inclined perturbation which forms an  angle $\theta$ with the direction of motion ${\bf i}$ (the $x$-direction).

The system for these perturbations is
\begin{eqnarray}\label{Couette-gen2}
	\left\{ \begin{array}{l}
		u'_t = -  {\bf u}\!\cdot\!\nabla u'+ {\rm Re}^{-1} \Delta u' -  (f u'_x+f' \cos \theta \,  w)- \dfrac{\partial p}{\partial x'}\\[5pt]
		v'_t = -  {\bf u}\!\cdot\!\nabla v'+ {\rm Re}^{-1} \Delta v' -  f v'_x + f' \sin \theta \, w- \dfrac{\partial p}{\partial y'}\\[5pt]
		w_t = -  {\bf u}\!\cdot\!\nabla w+  {\rm Re}^{-1}\Delta w -  f w_x- \dfrac{\partial p}{\partial z}\\	[5pt]	
		\dfrac{\partial u'}{\partial x'}+\dfrac{\partial v'}{\partial y'}+\dfrac{\partial w}{\partial z} =0 ,\\
	\end{array}  \right.
\end{eqnarray}
where
\be
\left\{\begin{array}{l}
	u'= \cos \theta  \, u + \sin \theta  \, v \\[2mm]
	v'= - \sin \theta  \, u + \cos \theta  \, v,
\end{array}\right. 
\ee
and
\be\label{xytran}
\left\{\begin{array}{l}
	x'= \cos \theta  \, x + \sin \theta  \, y \\[2mm]
	y'= - \sin \theta  \, x + \cos \theta  \, y.
\end{array}\right. 
\ee

We note that when $\theta \to 0$ then $x'\to x$, $y'\to y$, $u'\to u$, $v'\to v$.

Now we consider \textit{tilted (stream) perturbations in the $x'$-direction}, i.e, those with 	$\dfrac{\partial }{\partial x'}\equiv 0$. The first equation of  (\ref{Couette-gen2}) becomes

\be
u'_t = -  {\bf u}\!\cdot\!\nabla u'+ {\rm Re}^{-1} \Delta u' -  (f u'_x+f' \cos \theta \,  w) .
\ee

We have the energy equation

\be \label{eneqvw2-0}\dfrac{d}{dt} [\dfrac{\beta \Vert u'\Vert^2}{2}]= -\beta (f' \cos\theta u',w) - \beta {\rm Re}^{-1}\Vert \nabla u' \Vert^2, \ee
where $\beta$ is an arbitrary  positive number to be chosen. 

Moreover, we have

\be\label{eneqvw2-1} \dfrac{d}{dt} [\dfrac{\Vert v'\Vert^2+ \Vert w\Vert^2}{2}]=  - {\rm Re}^{-1} (\Vert \nabla v' \Vert^2+ \Vert \nabla w )\Vert^2+ (f' \sin \theta \, v', w).\ee

We note that as $\theta \to 0$ the energy equations tend to the energy equations obtained for the streamwise perturbations. Moreover, in the case $\dfrac{\partial }{\partial x'}\equiv 0$,  if $\theta \to \dfrac{\pi}{2}$, since $y'\to -x$,  $v'\to -u$, $x'\to y$, 
$\dfrac{\partial v'}{\partial y'} \to \dfrac{\partial u}{\partial x}$, the energy equation (\ref{eneqvw2-1}) becomes that for spanwise perturbations \cite{Falsaperla.Giacobbe.Mulone.2019}.

Defining 

\be H =\dfrac{1}{2}[\Vert v' \Vert^2 + \Vert w \Vert^2 ], \ee
we have
\be\label{stimV}
\begin{array}{l}
	\dot H= (f' \sin \theta \, v', w) - {\rm Re}^{-1} [\Vert \nabla v' \Vert^2+\Vert \nabla w \Vert^2] \le \\[3mm]
	\le \left( \dfrac{1}{{\rm Re}_c} - \dfrac{1}{{\rm Re}}\right)[\Vert \nabla v' \Vert^2+\Vert \nabla w \Vert^2],
\end{array}\ee 
where 
\be\label{maxRe}\dfrac{1}{{\rm Re}_c}= \max_{{\cal S}_0} \dfrac{(f' \sin \theta \, v',w)}{\Vert \nabla v'\Vert^2+\Vert \nabla w \Vert^2}, \ee
and ${\cal S}_0$ is the space of the \textit{physically admissible fields }
\be \begin{array}{l}\label{spaceS}
	{\cal S}_0= \{v', w \in H^1 (\Omega), \; v'=w=0 \hbox{ on the boundaries},\\[3mm] \q v'_{y'}+w_z=0,\q  \Vert \nabla v' \Vert+ \Vert \nabla w \Vert>0\}.
\end{array}
\ee

Falsaperla et al. \cite{Falsaperla.Giacobbe.Mulone.2019} proved that, for a fixed wavelength $\lambda= \dfrac{2\pi}{a}$ and a given angle $\theta$, one has

\be\label{fin1}{\rm Re}_c={\rm Re}_{Orr}(\frac{2\pi}{\lambda\sin \theta})/\sin \theta,\ee
where ${\rm Re}_{Orr}(\frac{2\pi}{\lambda\sin \theta})$ is the critical Reynolds number for given wavelength $\lambda$ and angle $\theta$.

In particular we note that for $\theta \to \frac{\pi}{2}$ we obtain the critical Orr Reynolds number for spanwise perturbations. For $\theta \to 0$ we obtain that the critical Reynolds number tends to $+ \infty$. Falsaperla et al. \cite{Falsaperla.Giacobbe.Mulone.2019} also proved that $\Vert u'\Vert \to 0$ as $t\to \infty$.

Formula \eqref{fin1} gives the critical value for a fixed positive wavelength $\lambda$. The minimum with respect to $\lambda$ in $(0, +\infty)$ is the nonlinear critical Reynolds number for tilted perturbations of an angle $\theta$:
\be\label{fin2}{\rm Re}_c = \min_{\lambda>0}{\rm Re}_{Orr}(\frac{2\pi}{\lambda\sin \theta})/\sin \theta={\rm Re}_{Orr}/\sin \theta,\ee where ${\rm Re}_{Orr}= 44.3$ for plane Couette flow and  ${\rm Re}_{Orr}= 87.6$ for plane Poiseuille flow. These results are shown in Figure 3 for Couette case.

The meaning of \eqref{fin2} is that when the Reynolds number is less than ${\rm Re}_c$ the linear and nonlinear energy of tilted perturbations of an angle $\theta$  and arbitrary wavelength $\lambda$ decay exponentially fast as $t\to +\infty$. For instance, if we consider tilted perturbations of an angle $\theta= 25^\circ$, the critical energy Reynolds number is ${\rm Re}_c= 104.82$ for plane Couette flow and ${\rm Re}_c= 207.51$ for plane  Poiseuille flow, for any $\lambda$. These values though lower than the experimental or numerical values however are bigger than the classical ones.

In order to compare the critical Reynolds numbers with the experiments and the numerical simulations, Falsaperla et al. \cite{Falsaperla.Giacobbe.Mulone.2019} considered \eqref{fin1} with fixed wavelengths given by the experiments of Prigent et al. \cite{Prigent.Gregoire.Chate.Dauchot.2003} and numerical calculations \cite{Barkley.Tuckerman.2007}. For instance, for 

i) $\theta=25^\circ$, $\lambda= 46$, experimental Reynolds number is about $395$, they obtained  approximately ${\rm Re}_{Orr}(\frac{2\pi}{46\sin 25^\circ})/\sin 25^\circ=369$,

ii)  $\theta=26^\circ$, $\lambda= 48$, experimental Reynolds number is about $385$, they obtained approximately ${\rm Re}_{Orr}(\frac{2\pi}{48\sin 26^\circ})/\sin 26^\circ=383$ (for other values see \cite{Falsaperla.Giacobbe.Mulone.2019}).

Therefore, they obtained that the energy stability limit are in a very good agreement with the experiments. This gives a possible solution to the Couette-Sommerfeld paradox.

\begin{figure}[h]{\label{tilted2}}
	\begin{center}
		\includegraphics[width=10cm]{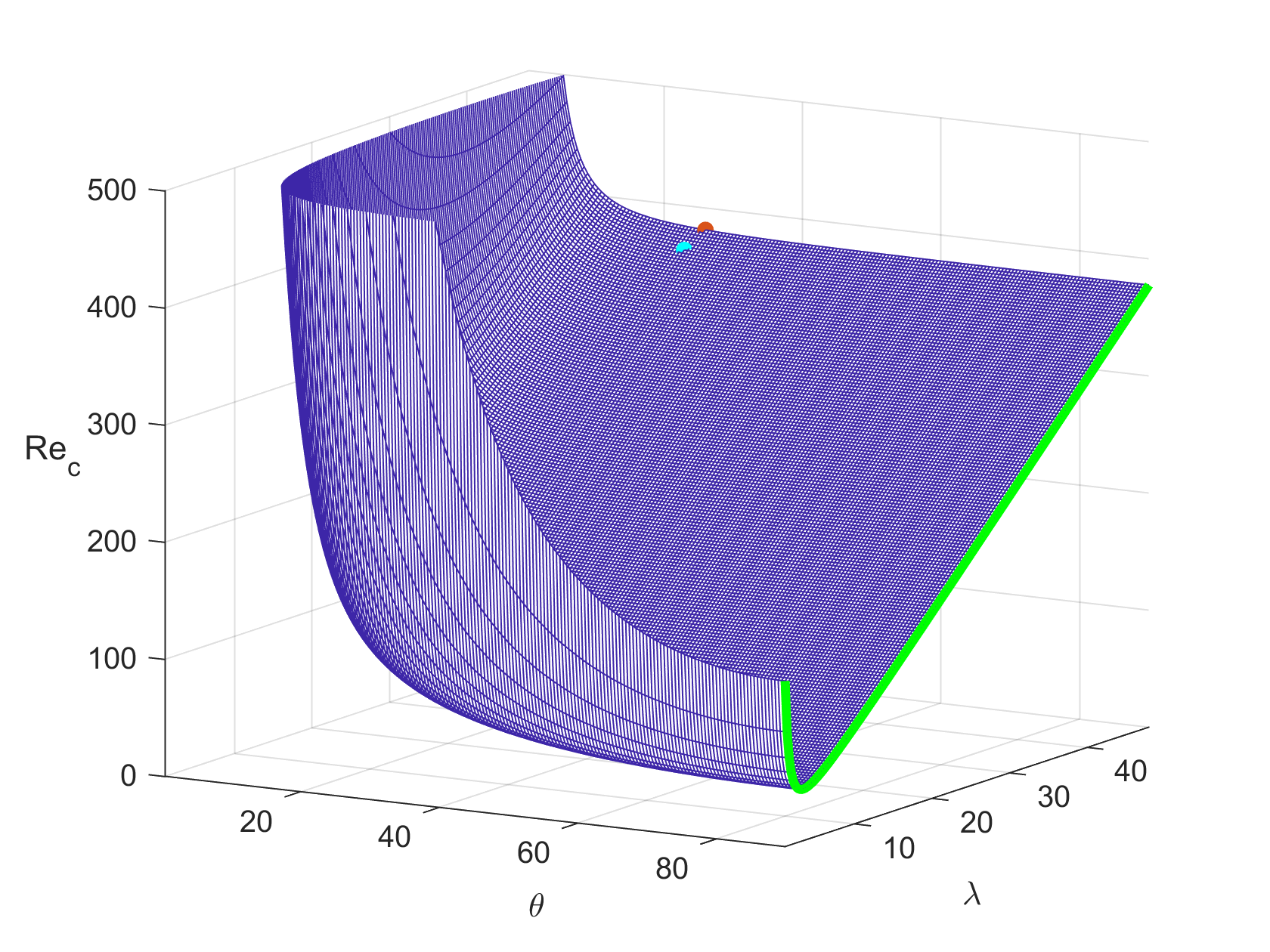}
	\end{center}  \caption{Plane Couette tilted energy Orr-Reynolds  number ${\rm Re}=\rm R_c$ as function of the wavelength $\lambda$ and angle $\theta$. The two points light blue and red  correspond to the cases i) and ii).  For Poiseuille flow a similar picture can be drawn.}
\end{figure}

\section{Conclusion}

We have proved sufficient conditions of global nonlinear stability of plane Couette and Poiseuille flows for any Reynolds number less than the ordinary limit of spanwise perturbations. For streamwise perturbations we have ri\-gorously proved nonlinear $L_2$-energy stability results for any Reynolds number and we have suggested a way to overcome an obvious contradiction with classical numerical results. 
We have introduced a space of physical perturbations for the maximum problem  and we have proved that the least stabilizing perturbations are two-dimensional (spanwise perturbations). This result justifies the previous study of Falsaperla et al. \cite{Falsaperla.Giacobbe.Mulone.2019} on the stability of tilted rolls  with axes in the $Oxy$-plane, and gives a possible solution of the Couette-Sommerfeld paradox.

We finally note that some similar results hold for shear flows in magneto-hydro\-dynamics (forthcoming paper). 

\vskip .3cm
{\bf Acknowledgments.}\ 
The research that led to the present paper was partially supported by the following Grants: 2017YBKNCE of national project PRIN of Italian Ministry for University and Research, PTR DMI-53722122113 and "ASDeA" of the University of Catania. The authors acknowledge also support from the project PONSCN 00451 CLARA-CLoud plAtform and smart underground imaging for natural Risk Assessment, SmartCities and Communities and Social Innovation. 
We also thank the group GNFM of INdAM for financial support.

\phantomsection
\vspace*{0.5cm}
{\footnotesize\begin{tabular}{rl}
		\hline
		& \\
		\label{A}$^{a}$
		& Universit{\`{a}} degli Studi di Catania\\
		& Dipartimento di Matematica e Informatica\\
		& Viale A. Doria 6\\
		& 9512500 Catania, Italy\\
		& E-mail: 
		\href{mailto:falsaperla@dmi.unict.it}{falsaperla@dmi.unict.it},
		\href{mailto:giuseppe.mulone@unict.it}{giuseppe.mulone@unict.it} \href{mailto:carla.perrone@unipa.it}{carla.perrone@unipa.it}\\
\end{tabular}}

\end{document}